# Prevalence of the adiabatic exchange-correlation potential approximation in time-dependent density functional theory

Roi Baer
*Institute of Chemistry and the Fritz Haber Center for Molecular Dynamics, The Hebrew University of Jerusalem, Jerusalem 91904 Israel.*

Time-dependent (TD) density functional theory (TDDFT) promises a numerically tractable account of many-body electron dynamics provided good simple approximations are developed for the exchange-correlation (XC) potential functional (XCPF). The theory is usually applied within the adiabatic XCPF approximation, appropriate for slowly varying TD driving fields. As the frequency and strength of these fields grows, it is widely held that memory effects kick in and the eligibility of the adiabatic XCPF approximation deteriorates irreversibly. We point out however that when a finite system of electrons in its ground-state is gradually exposed to a very a high-frequency and eventually ultra-strong homogeneous electric field, the adiabatic XCPF approximation is in fact rigorously applicable. This result not only helps to explain recent numerical results for a 1D-helium atom subject to a strong linearly-polarized laser pulse (Thiel et al, Phys. Rev. Lett. 100, 153004, (2008)) but also shows that it is applicable to any number of electrons and in full 3D.

Time-dependent density functional theory (TDDFT) is an in-principle exact approach to the quantum dynamics of electrons under time-dependent fields[1, 2]. So far, the theory is very successful in many cases where the adiabatic-linear-response limit holds. These are small excitations over the ground state and the theory draws upon the ground-state density functional theory[3] and in particular the static exchange-correlation potential functional (XCPF)[4]. In strong fields and faster processes the time-dependent exchange-correlation potential at time $t$ should depend also on the density at previous times $t' < t$, an effect referred to as "memory"[5]. It is then expected that "memory effects" must dominate or at least be important and cannot be neglected. Even in linear response memory effects may be important, especially when special correlation effects are dominant[6]. While there have been recently several works on the deployment of memory with TDDFT or time-dependent current-density functional theory (TDCDFT) [5, 7-16] there is not yet available reliable and generally satisfying exchange-correlation potential functionals with memory. As part of the effort to develop new approaches, relatively simple model systems have been used to reveal some of the properties of the exact time-dependent XC potential.[17-19] A somewhat surprising result emerged from these studies: memory effects were seen to be small or even negligible in a certain range of fast and non-perturbative cases.

The purpose of this letter is to show that the adiabatic approximation is in fact of much greater generality and has a broader range of applicability than previously suspected. We show that for electrons in finite (molecular) systems, evolving in time under the influence of a high-frequency strong homogeneous electric-field $\vec{E}(t)$ the adiabatic approximation is valid. The oscillating electric field takes the following form: $\vec{E}(t) = \omega^2 X(t)(p_x \cos\omega t, p_y \sin\omega t, 0)$, where $\omega$ is the frequency, $\vec{p}$ is a polarization vector (assumed in the x-y plane for simplicity). We assume that the field is zero for negative times, so is turned on slowly starting at $t = 0$, as described by the ramp envelope $\omega^2 X(t)$. The field $\vec{E}$ is usually applied as a strong laser pulse propagating in the z direction, in which case one must assume eligibility of the dipole approximation.

The electrons in the molecular system start from their ground state $\psi_{gs}(\{\vec{r}\})$, where $\{\vec{r}\}$ is shorthand notation for the position and spin coordinates of all electrons in the system. The Schrödinger equation is $i\dot{\psi}(t;\{\vec{r}\}) = (\hat{F} + \hat{V}(t))\psi(t,\{\vec{r}\})$ where $\hat{F} = \sum_j \left(-\frac{1}{2}\nabla_j^2\right) + \sum_{k\neq j} \frac{1}{|\vec{r}_j - \vec{r}_k|}$ is the sum of kinetic energy and potential energy of electron-electron repulsion. The external potential can be written as: $\hat{V}(t) = \int v(\vec{r},t)\hat{n}(\vec{r})d^3r$, where $\hat{n}(\vec{r}) = \sum_j \delta(\vec{r} - \vec{r}_j)$ is the number-density operator and:

$$v(\vec{r},t) = v_N(\vec{r}) + \vec{E}(t) \cdot \vec{r}, \qquad (1)$$

with $v_N(\vec{r})$ the potential of the external force on the electrons (originating from the static nuclei, for example). The TDDFT procedure replaces the interacting electron system by a "non-interacting" one, namely the wavefunction is replaced by a time-dependent Slater determinant that evolves in time starting from the Kohn-Sham ground-state determinant. The basic relation between the interacting and non-interacting systems is that both have the same density $n(\vec{r},t)$ for all times. This serves to define uniquely the potential exerted on the non-interacting system which is written as[1]:

$$v_s(\vec{r},t) = v(\vec{r},t) + v_H[n(t)](\vec{r}) + v_{XC}[n](\vec{r},t), \qquad (2)$$

where the $v_H[n(t)](\vec{r}) = \int n(\vec{r}',t)/|\vec{r}-\vec{r}'|d^3r$ is the Hartree potential and $v_{XC}[n](\vec{r},t)$ is the universal, exact yet forever elusive XCPF. The central issue in TDDFT is the developments of useful approximations for the XCPF. The adiabatic approximation is extremely popular and consists of simply taking the ground-state XCPF and plugging in the instantaneous density (assuming v-representability):

$$v_{XC}^{ad}[n](\vec{r},t) = v_{XC}^{gs}[n(t)](\vec{r}) \quad (3)$$

Can the drastic approximation in Eq. (3) be expected to work exactly for high frequencies and strong fields? One case where it certainly does is that of electrons in a harmonic trap ("Hooke's atom"). This can be readily proved using the Harmonic potential theorem [20] and Galilean covariance of the exact exchange correlation potential functional [21]. But the harmonic potential is a very special case and the question lingers for more general circumstances.

In order to set the stage for our argument, we move to the acceleration-frame, in which a *free electron* subject to the given electric field is at rest. If $\vec{x}(t)$ is the trajectory of such an electron in the rest-frame, then $\ddot{\vec{x}}(t) = -E(t)$. The initial conditions of the trajectory will be chosen so as to have the accelerated observer not move *on the average*, thus making the trajectory motion as purely vibrational as possible. Furthermore, we may redefine the origin so as to have $\vec{x}(0) = 0$. The so called Kramers-Henneberger (KH) gauge [22, 23] is the formulation of the Schrödinger equation in the acceleration frame. The Schrödinger equation in this gauge is:

$$i\dot{\psi}_{KH}(\{\vec{r}\},t) = [\hat{F} + \hat{V}_{KH}(t)]\psi_{KH}(\{\vec{r}\},t) \quad (4)$$

Where $\hat{V}_{KH} = \int v_{KH}(\vec{r},t)\hat{n}(\vec{r})d^3r$ and $v_{KH}(\vec{r},t) = v_N(\vec{r}+\vec{x}(t))$. The wave functions in the rest- and accelerated-frames relate through:

$$\psi(\{\vec{r}\},t) = e^{i\theta(t)}\psi_{KH}(\{(\vec{r}-\vec{x}(t))\},t) \quad (5)$$

Where $\theta(t)$ is a time-dependent phase. We may take $\psi_{KH}(\{\vec{r}\},0) = \psi_{gs}(\{\vec{r}\})$ as an initial condition just as in the rest-frame. This entire procedure is exact as only a change of frame of reference was made. We note that the expectation value of the electron number-density in the rest-frame $n(\vec{r},t)$ is related to that seen in the accelerated-frame by:

$$n(\vec{r},t) = n_{KH}(\vec{r}-\vec{x}(t),t) \quad (6)$$

We now wish to discuss the case of a high-frequency electric field. In particular we consider a finite $\omega$ and make an approximation which becomes exact as $\omega$ is increased to infinity. The field envelope $X(t)$ changes slowly on the time scale of $2\pi\omega^{-1}$, thus the Newtonian trajectory can be approximated by:

$$\vec{x}(t) = X(t)(p_x\cos\omega t, p_y\sin\omega t, 0) \quad (7)$$

(One may verify this by differentiation, neglecting terms of order $\omega^{-1}\dot{X}/X$). This approximation becomes exact when $\omega \to \infty$ while $X(t)$ is left unchanged. It is clear that in this limit the electric field $\vec{E}$ which is proportional to $\omega^2 X$ grows to infinity as well. Thus we are now looking at the *high-frequency strong-field limit*.

The next step is to introduce a "slow" time variable $t'$ in addition to the fast "time" $t$. This allows for a Born-Oppenheimer-Floquet approach. The details are discussed in ref. [24] where in the high frequency limit, it is shown that the relevant Schrödinger equation is:

$$i\frac{\partial}{\partial t'}\psi_{KH}(\{\vec{r}\},t') = (\hat{F} + \hat{V}_0(t'))\psi_{KH}(\{\vec{r}\},t'), \quad (8)$$

where $\hat{V}_0(t') = \int v_0(\vec{r},t')\hat{n}(\vec{r})d^3r$ and:

$$v_0(\vec{r},t') = \frac{1}{2\pi}\int_0^{2\pi} v_N(\vec{r}+X(t')(p_x\cos\theta, p_y\sin\theta, 0))d\theta. \quad (9)$$

Clearly, $v_0(\vec{r},t)$ is the fast-time-average of the KH potential $v_{KH}(\vec{r},t)$.

We ended up with a Schrödinger equation (Eq. (8)) having a new "averaged" potential without *high-frequency*. This fact can be now used to show that the adiabatic functional of Eq. (3) is applicable to this problem. Indeed, consider the limit of very slowly varying envelop $X(t)$ (henceforth we replace $t'$ by $t$ for simpler notation). According to the adiabatic theorem $\psi_{KH}(\vec{r},t)$ is (up to a time-dependent phase) the instantaneous ground state of the averaged KH Hamiltonian $\hat{H}_0(t) = \hat{F} + \hat{V}_0(t)$. This instantaneous ground-state density $n_{KH}(\vec{r},t)$ is uniquely mapped to the KH potential $v_0(\vec{r},t)$, as follows from the Hohenberg-Kohn theorem [3]. Thus, the usual machinery of the Kohn-Sham method[4] can be used to determine the time dependent density $n_{KH}(\vec{r},t)$ by erecting a system of non-interacting electrons which in their ground state have their determinantal wave function spanned by spin-orbitals $\varphi_j(\vec{r},t)$ and density $n_{KH}(\vec{r},t) = \sum_j |\varphi_j(\vec{r},t)|^2$. These non-interacting electrons are subject to an external force derived from the potential:

$$v_{s,KH}(\vec{r},t) = v_0(\vec{r},t) + v_H[n_{KH}(t)](\vec{r}) + v_{XC}^{gs}[n_{KH}(t)](\vec{r}) \quad (10)$$

Where $v_H[n](\vec{r}) = \int n(\vec{r}')/|\vec{r}-\vec{r}'|d^3r$ is the Hartree potential. Now, when the ramp is not infinitely slowly changing, the adiabatic approximation may still be useful. Indeed, this is the limit where it is expected to hold. Basically, the approximation consists in solving the TD Kohn-Sham equations:

$$i\dot{\varphi}_j(\vec{r},t) = \left[-\frac{1}{2}\nabla^2 + v_{s,KH}(\vec{r},t)\right]\varphi_j(\vec{r},t). \quad (11)$$

Equations (10) and (11) constitute an adiabatic TDDFT approach to the time-dependent electron density in the KH frame. They are expected to be valid because the underlying Hamiltonian is slowly changing.

Once the $n_{HK}(\vec{r},t)$ is determined, one can, if so needed, return to the rest frame density $n(\vec{r},t)$ via Eq. (6). Because adiabatic functionals are Galilean covariant[21] the XC potential in the rest frame is simply $v_{XC}^{gs}[n(t)](\vec{r},t)$, showing that we may have stayed in the original frame and used the adiabatic approximation to begin with.

Summarizing, we have shown that the adiabatic XCPF approximation is valid not only in the static limit but also on the opposite extreme, when a highly oscillatory super-strong electric-field is operative. The suitability of the adiabatic approach for high-frequency and high-field cases has been seen numerically in 1-dimensional 2-electron system.[18, 19] The development brought here explains these results and also extends them to any number of electrons and to 3 dimensions, including all polarizations.

At first sight, the present result seems paradoxical in view of the well-known fact that the XC kernel $f_{XC}(q,\omega)$ of the homogeneous electron gas (HEG) at high frequency ($\omega \to \infty$) and low wave vector $q \to 0$ is different from that of low frequency ($\omega \to 0$) which corresponds to the ground state kernel[25]. This apparent discrepancy is probably due to the infinite nature of the HEG which precludes application of the Kramers-Henneberger gauge as done for finite systems.

Another interesting question remains unanswered: does this conclusion hold for other types of high frequency perturbations? One case, of physical significance, is in high-frequency laser pulses, for which the dipole-approximation may cease to be justified, since the wave length drops when the frequency increases (however, pretty high frequencies can be reached before the dipole approximation ceases to be eligible).

While we are not able to answer all questions at present, it is very comforting to find that the adiabatic XCPF approximation has an exact high-frequency strong-field limit. This fact may assist in developing new approaches, based on the adiabatic approximation for strong fields and intermediate frequencies. Furthermore, we find that improvement of approximate ground state XCPFs is an important step towards improved TDXCPFs.

This research was supported by the Israel Science Foundation (grant number 962/06).